\documentclass[aps,pra,showpacs,superscriptaddress,preprint]{revtex4}
\usepackage{graphicx}

\catcode`ð=\active
 \defð{\u{g}}
 \catcode`Ð=\active
 \defÐ{\u{G}}
 \catcode`Ý=\active
 \defÝ{\. I}
 \catcode`ö=\active
 \defö{\"{o}}
 \catcode`Ö=\active
 \defÖ{\"O}
 \catcode`ü=\active
 \defü{\"{u}}
 \catcode`Ü=\active
 \defÜ{\"{U}}
 \catcode`Þ=\active
 \defÞ{\c{S}}
 \catcode`þ=\active
 \defþ{\c{s}}
 \catcode`ý=\active
 \defý{{\i}}
 \catcode`ç=\active
 \defç{\d{c}}
 \catcode`Ç=\active
 \defÇ{\d{C}}

\begin{document}

\title{Approximate $\ell$-State Solutions of the Klein-Gordon Equation for Modified Woods-Saxon
Potential With Position Dependent Mass }

\author{\small Altuð Arda}
\email[E-mail: ]{arda@hacettepe.edu.tr}\affiliation{Department of
Physics Education, Hacettepe University, 06800, Ankara,Turkey}
\author{Ramazan Sever}
\email[E-mail: ]{sever@metu.edu.tr}\affiliation{Department of
Physics, Middle East Technical University, 06800, Ankara,Turkey}

\date{\today}

\begin{abstract}
The radial part of the Klein-Gordon equation for the generalized
Woods-Saxon potential is solved by using the Nikiforov-Uvarov
method in the case of spatially dependent mass within the new
approximation scheme to the centrifugal potential term. The energy
eigenvalues and corresponding normalized eigenfunctions are
computed. The solutions in the case of constant mass are
also studied to check out the consistency of our new approximation scheme.\\
Keywords: Woods-Saxon Potential, Position Dependent Mass,
Klein-Gordon Equation, Nikiforov-Uvarov Method
\end{abstract}
\pacs{03.65.Fd, 03.65.Ge}

\maketitle

\newpage

\section{Introduction}
The investigation of the quantum mechanical systems in the case of
position dependent mass (PDM) following of works by v. Roos, and
Levy-Leblond [1, 2] have recently been received great attentions.
This is so because such solutions are aviable in wide range of
different areas, for example, in the study of impurities in
crystals [3-5], or of electronic properties of quantum wells, and
quantum dots [6], and in semiconductor heterostructures [7].
Yahiaowi, and Bentaiba [8] have studied the
weak-pseudo-Hermiticity in the case of PDM, Ganguly, and Nieto
have extended the second-order supersymmetric approach to the
systems with coordinate dependence mass [9]. In Ref. [10], some
new shape-invariant, exactly solvable potentials are generated by
using a specific ansatz in the point of PDM-case. Ju et al. have
been studied the dynamics of a quasi-free particle in an effective
potential arising from the dependence of the mass on coordinates,
and analyzed the eigenfunctions and probability  densities for
s-waves [11].

Another interesting area received a lot of attentions is that
solving the Schr\"{o}dinger (SE), and Dirac equations in the case
of PDM. To solve the above equations has been used different
methods, and approaches for different potentials, such as deformed
algebras in Coulomb potential [12], in the content of
supersymmetric quantum mechanics [13, 14, 15, 16], quadratic
algebra approach [17], numerical analysis of a square potential by
using appropriate matching conditions [18], point canonical
transformation applying on harmonic oscillator, Coulomb and Morse
class of potentials [19], finding the non-relativistic Green's
functions with PDM for harmonic oscillator [20], Coulomb potential
in Dirac equation [21], Morse potential in PDM background [22], a
series solution of the SE for Cornell potential [23], finding the
bound states of Rosen-Morse and Scarf potentials via general point
canonical transformation [24].

In the present work, we give the approximate solutions, and
corresponding wave functions of the radial Klein-Gordon (KG)
equation for the Woods-Saxon (WS) potential in the case of PDM. We
investigate the energy spectrum, and the corresponding
eigenfunctions of the generalized WS potential by using a new
approximation to the centrifugal potential. In order to find the
spectrum we use the NU-method in the case of exponentially mass
distribution varying with coordinate. The NU-method is a powerful
tool to solve of the second order linear differential equations
with special orthogonal functions. In this method, the
differential equation is turned into a hypergeometric type
equation by using a transformation on coordinate [25].

The organization of this work is as follows. In Section II, we
solve the radial part of the KG-equation for generalized WS
potential by using the NU-method within the framework of an
approximation to the centrifugal term. We compute the energy
eigenvalues and corresponding eigenfunctions, and also give the
results for the case of the constant mass to control the
consistency of our new approximation. We write our conclusions in
Section III.
\section{Nikiforov-Uvarov Method and Calculations}
In spherical coordinates, the radial part of the Klein-Gordon
equation can be written as [30]

\begin{eqnarray}
\Big\{\frac{\hbar^2}{2m}\,\frac{d^2}{dr^2}-\frac{\hbar^2\ell(\ell+1)}{2mr^2}-\frac{1}{2mc^2}
[m^2c^{4}-(E-V(r))^2]\Big\}\phi(r)=0\,,
\end{eqnarray}
where $\ell$ is the angular-momentum quantum number, $E$ is the
energy of the particle, $m$ is the rest mass, and $c$ is the
velocity of the light.

The generalized WS potential can be written of the form [31]

\begin{eqnarray}
V(r)=\,-\,\frac{V_0}{1+qe^{\beta(r-r_0)}}\,,\,\,\,\,\,(0 \leq r
\leq \infty)\,.
\end{eqnarray}
where $V_0$ is the potential depth, $\beta$ is a short notation,
i.e. $\beta \equiv 1/a$, $a$ is diffuseness of the nuclear
surface, $r$ is the center-of-mass distance between the projectile
and target nucleus, and $r_0$ is the width of the potential, which
is proportional with target mass number $A$. $q$ is the
deformation parameter, and arbitrarily taken to be a real
constant. The WS potential is widely used in the coupled-channels
calculations in heavy-ion physics. This model explains the
single-particle motion during a heavy-ion collisions [26-29].

Let us write the potential as

\begin{eqnarray}
V(x)=\,-\,\frac{V_0}{1+qe^{\beta x}}\,,
\end{eqnarray}
where $x=(r-r_0)$. Eq. (1) can not be solved exactly because of
the centrifugal potential term for $\ell \neq 0$\,. The nuclear
distance $r$ can not fluctuate very far from the equilibrium for
rather high vibrational levels [32], which gives small $x$-values.
So the centrifugal potential term can be expand about $x=0$ as the
following

\begin{eqnarray}
V_{1}(r)=\,\frac{\ell(\ell+1)}{r^2}\,=\,\frac{D}{(1+\,\frac{x}{r_{0}}\,)^2}=D(1-2
\,(\frac{x}{r_{0}})\,
+3\,(\frac{x}{r_{0}})^2+ \ldots)\,,
\end{eqnarray}
where the parameter $D$ in the above equation is given as
$D=\frac{\hbar^2\ell(\ell+1)}{2mr^2_{0}}$\,.

Instead, we suggest to replace $V_1(r)$ by the following potential
form [33]

\begin{eqnarray}
V'_{1}(x)=(1+qe^{\beta x})^{-2}\Big[DD_{0}(1+qe^{\beta
x})^2+DD_{1}(1+qe^{\beta x})+DD_{2}\Big]\,,
\end{eqnarray}
where the parameters $D_{0}, D_{1}$\,, and $D_{2}$ are arbitrary
constants.

Expanding the potential $V'_{1}(x)$ around $x=0$ under the same
condition, and than combining equal powers with Eq. (4), one can
find the arbitrary constants $D_i (i=0, 1, 2)$ in the new form of
the potential as

\begin{eqnarray}
D_{0}&=&1-\frac{(1+q)^2}{\beta r_{0}q^2}\Bigg[-\frac{3}{\beta r_{0}}+1\Bigg]\,,\\
D_{1}&=&\frac{(1+q)^2}{\beta r_{0}q^2}\Bigg[-\frac{6(1+q)}{\beta r_{0}}+3q-1\Bigg]\,,\\
D_2&=&\frac{(1+q)^3}{\beta r_0q^2}\Bigg[\frac{3(1+q)}{\beta
r_0}+\frac{1-q}{2}\Bigg]\,.
\end{eqnarray}
where it can be seen that the new parameters $D_{0}, D_{1}$\,, and
$D_{2}$ are real, dimensionless parameters, and dependent to the
numerical values of the quantum system under consideration.

On the other hand, we prefer to use the following position
dependent mass function

\begin{eqnarray}
m(x)=\,m_0\,\Big[1-\,\frac{m_1}{m_0}\,\Big(1+qe^{\beta x
}\Big)^{-1}\Big]\,\,\,(m_0>m_1)\,,
\end{eqnarray}
where $m_0$ and $m_1$ are two arbitrary positive parameters. The
mass function is finite at infinity, and enables us to solve
analytically the KG-equation given by Eq. (1), and to check out
the limit of the case of the constant mass.

Substituting Eq. (5), and (9) into Eq. (1), we get

\begin{eqnarray}
\Big\{\frac{d^2}{dx^2}&-&\Big[\,\frac{1}{\hbar^2c^2}(m^2_{0}c^4-E^2)+\frac{\hbar^2\ell(\ell+1)}{r^2_{0}}
D_{0}\Big]\nonumber\\&+&\Big[\frac{2}{\hbar^2c^2}(EV_{0}+m_{0}m_{1}c^4)-\frac{\hbar^2\ell(\ell+1)}{r^2_{0}}
D_{1}\Big]\frac{1}{1+qe^{\beta
x}}\nonumber\\&+&\Big[\frac{1}{\hbar^2c^2}(V^2_{0}-m^2_{1}c^4)-\frac{\hbar^2\ell(\ell+1)}{r^2_{0}}
D_2\Big]\frac{1}{(1+qe^{\beta
x})^2}\Big\}\phi(x)=0\,
\end{eqnarray}

By using the transformation $z=2(1+qe^{\beta x})^{-1}$, we have

\begin{eqnarray}
\frac{d^2\phi(z)}{dz^2}\,+\,\frac{2(1-z)}{z(2-z)}\,\frac{d\phi(z)}{dz}\,+\,\frac{1}{[z(2-z)]^2}
\left[-a_1^2z^2-a_2^2z-a_3^2\right]\phi(z)=0\,.
\end{eqnarray}
where

\begin{eqnarray}
a^2_{1}&=&\omega^2_{1}\ell(\ell+1)D_{2}+\omega^2_{2}(m^2_{1}c^4-V^2_{0})\,,\nonumber\\
a^2_{2}&=&2[\omega^2_{1}\ell(\ell+1)D_{1}-2\omega^2_{2}(m_{0}m_{1}c^4+EV_{0})]\,,\nonumber\\
a^2_{3}&=&4[\omega^2_{1}\ell(\ell+1)D_{0}+\omega^2_{2}(m^2_{0}c^4-E^2)]\,.
\end{eqnarray}
and $\omega^2_{1}=1/\beta^2r^2_{0}$\,, and
$\omega^2_{2}=1/\hbar^2c^2\beta^2$\,.

To apply the NU-method, we rewrite Eq. (11) in the following form

\begin{eqnarray}
\phi^{\prime\prime}(z)+\,\frac{\tilde{\tau}(z)}{\sigma(z)}\,\phi^{\prime}(z)
+\,\frac{\tilde{\sigma}(z)}{\sigma^2(z)}\,\phi(z)=0,
\end{eqnarray}
where $\sigma(z)$ and $\tilde{\sigma}(z)$ are polynomials with
second-degree, at most, and $\tilde{\tau}(z)$ is a polynomial with
first-degree. By using the following transformation for the total
wave function

\begin{eqnarray}
\phi(z)=\xi(z)\psi(z)
\end{eqnarray}
we get a hypergeometric type equation

\begin{eqnarray}
\sigma(z)\psi^{\prime\prime}(z)+\tau(z)\psi^{\prime}(z)+\lambda\psi(z)=0,
\end{eqnarray}
where $\xi(z)$ satisfies the equation

\begin{eqnarray}
\xi^{\prime}(z)/\xi(z)=\pi(z)/\sigma(z).
\end{eqnarray}
and the other part, $\psi(z)$, is the hypergeometric type function
whose polynomial solutions are given by

\begin{eqnarray}
\psi_n(z)=
\,\frac{b_n}{\rho(z)}\,\frac{d^n}{dz^n}[\sigma^n(z)\rho(z)],
\end{eqnarray}
where $b_n$ is a normalization constant, and the weight function
$\rho(z)$ must satisfy the condition

\begin{eqnarray}
\frac{d}{dz}[\sigma(z)\rho(z)]=\tau(z)\rho(z).
\end{eqnarray}
The function $\pi(z)$ and the parameter $\lambda$ required for
this method are defined as follows

\begin{eqnarray}
\pi(z)=\,\frac{\sigma^{\prime}(z)-\tilde{\tau}(z)}{2}\,
\pm\,\sqrt{(\frac{\sigma^{\prime}(z)-\tilde{\tau}(z)}{2})^2-\tilde{\sigma}(z)+k\sigma(z)}\,,
\end{eqnarray}
\begin{eqnarray}
\lambda=k+\pi^{\prime}(z)
\end{eqnarray}
The constant $k$ is determined by imposing a condition such that
the discriminant under the square root should be zero. Thus one
gets a new eigenvalue equation

\begin{eqnarray}
\lambda&=&\lambda_n=-n\tau^{\prime}-\,\frac{n(n-1)}{2}\,\sigma^{\prime\prime}\,,
(n=0, 1, 2, \ldots)
\end{eqnarray}
where

\begin{eqnarray}
\tau(z)=\tilde{\tau}(z)+2\pi(z)\,.
\end{eqnarray}
and the derivative of $\tau(z)$ must be negative.

Comparing Eq. (11) with Eq. (13), we have
\begin{eqnarray}
\tilde{\tau}(z)=2(1-z)\,,\,\,\,\,\,\sigma(z)=z(2-z)\,,\,\,\,\,\,
\tilde{\sigma}(z)=-a_1^2z^2-a_2^2z-a_3^2
\end{eqnarray}
Substituting this into Eq. (19), we get

\begin{eqnarray}
\pi(z)=\pm\sqrt{(a_1^2-k)z^2+(a_2^2+2k)z+a_3^2}.
\end{eqnarray}
The constant $k$ can be determined by the condition that the
discriminant of the expression under the square root has to be zero

\begin{eqnarray}
(a_2^2+2k)^2-4a_3^2(a_1^2-k)=0\,.
\end{eqnarray}

The roots of $k$ are
$k_{1,2}=\,-\,\frac{1}{2}\,a_2^2\,-\,\frac{1}{2}\,a_3^2\mp\,\frac{1}{2}\,a_3A$,
where $A=\sqrt{a_3^2+2a_2^2+4a_1^2}$. Substituting these values
into Eq.(19), we get for $\pi(z)$ for $k_1$

\begin{eqnarray}
\pi(z)=\mp
\Big[\Big(\,\frac{A}{2}\,-\,\frac{a_3}{2}\,\Big)z+a_3\Big]\,,
\end{eqnarray}
and for $k_2$

\begin{eqnarray}
\pi(z)=\mp
\Big[\Big(\,\frac{A}{2}\,+\,\frac{a_3}{2}\,\Big)z-a_3\Big]\,,
\end{eqnarray}

Now we find the polynomial $\tau(z)$ from $\pi(z)$ for the second
choice as

\begin{eqnarray}
\tau(z)=2+2a_3-2\Big(\,\frac{A}{2}\,+\,\frac{a_3}{2}\,+1\,\Big)z.
\end{eqnarray}
so its derivative
$-2\Big(\,\frac{A}{2}\,+\,\frac{a_3}{2}\,+1\,\Big)$ is negative.
We have from Eq. (20)

\begin{eqnarray}
\lambda=\,-\,\frac{1}{2}\,\Big(a_2^2+a^2_3+Aa_3+A+a_3\Big)\,,
\end{eqnarray}
and Eq. (21) gives us

\begin{eqnarray}
\lambda_n=2n\Big(\,\frac{A}{2}\,+\,\frac{a_3}{2}\,+1\,\Big)+n^2-n\,.
\end{eqnarray}
Substituting the values of the parameters given by Eq. (12), and
setting $\lambda=\lambda_n$, one can find the energy eigenvalues
for any $\ell$-states

\begin{eqnarray}
E_{n,\ell}&=&-\frac{V_{0}[4\omega^2_{2}V^2_{0}+N^2-4\omega^2_{2}\ell(\ell+1)(D_{1}+D_{2})
+4m_{1}c^4\omega^2_{2}(2m_{0}-m_{1})]}{2(N^2+4\omega^2_{2}V^2_{0})}\nonumber\\&\pm&\frac{N}{\omega_{2}}
\sqrt{\frac{\omega^2_{1}(2D_{0}+D_{1}+D_{2})\ell(\ell+1)+2m^2_{0}c^4\omega^2_{2}}
{2(N^2+4\omega^2_{2}V^2_{0})}-\Bigg[\frac{\omega^2_{1}(D_{1}+D_{2})\ell(\ell+1)}{N^2+4\omega^2_{2}V^2_{0}}\Bigg]^2
-\frac{1}{16}+\tilde{m}^2_{1}\,}\,,\nonumber\\
\end{eqnarray}
where the energy eigenvalues with (+) sign correspond to particle,
and the one with (-) sign correspond to antiparticle. Two
parameters in the above expression are

\begin{eqnarray}
\tilde{m}_{1}=\frac{\sqrt{8m_{1}c^4\omega^2_{2}(2m_{0}-m_{1})
[4\omega^2_{1}(D_{1}+D_{})\ell(\ell+1)-2\omega^2_{2}m_{1}c^4(2m_{0}-m_{1})-
(N^2+4\omega^2_{2}V^2_{0})]\,}}{N^2+4\omega^2_{2}V^2_{0}}\,,\nonumber\\
\end{eqnarray}
and
\begin{eqnarray}
N=-(2n+1)+\sqrt{1+4a^2_1}\,.
\end{eqnarray}
We see that the energy levels for particles and antiparticles are
symmetric, and the ground state energy is different from zero. We
summarize some numerical results in Table I to see the effect of
the spatially dependent mass parameter $m_{1}$ on the energy
eigenvalue of bound states. It is observed that the energy levels
are strongly dependent on the parameter, and the increase of the
energy eigenvalues in the existence of $m_{1}$ is very
significant. It has to be stress that the higher numerical values
of the parameter $m_{1}$ give positive values for the bound
states.

The energy spectra in the case of constant mass is obtained by
setting $m_{1}=0$ in Eq. (32) which gives us $\tilde{m}_1=0$\,,
and we get

\begin{eqnarray}
E^{m_{1}=0}_{n,\ell}&=&-\frac{V_{0}[4\omega^2_{2}V^2_{0}+N^2-4\omega^2_{2}\ell(\ell+1)(D_{1}+D_{2})]}
{2(N^2+4\omega^2_{2}V^2_{0})}\nonumber\\&\pm&\frac{N}{\omega_{2}}
\sqrt{\frac{\omega^2_{1}(2D_{0}+D_{1}+D_{2})\ell(\ell+1)+2m^2_{0}c^4\omega^2_{2}}
{2(N^2+4\omega^2_{2}V^2_{0})}-\Bigg[\frac{\omega^2_{1}(D_{1}+D_{2})\ell(\ell+1)}{N^2+4\omega^2_{2}V^2_{0}}\Bigg]^2
-\frac{1}{16}\,}\,,\nonumber\\
\end{eqnarray}
where
\begin{eqnarray}
N'=-(2n+1)+\sqrt{1+4a'^2_1}\,,\,\,\,a'^2_1=a^2_1(m_1 \rightarrow
0).
\end{eqnarray}

It is seen that the result for the case of constant mass is the
same with those obtained in Ref. (29).

In order to find the eigenfunctions, we first compute the weight
function from Eq. (18)

\begin{eqnarray}
\rho(z)=z^{a_3}(2-z)^{A}\,,
\end{eqnarray}
and the wave function becomes

\begin{eqnarray}
\psi_{n\ell}\,(z)&=&\,\frac{b_n}{z^{a_3}(2-z)^{A}}\,\frac{d^n}{dz^n}\,\left[
\,z^{n+a_3}\,(2-z)^{n+A}\right]\,.
\end{eqnarray}
where $b_n$ is a normalization constant. The polynomial solutions
can be written in terms of the Jacobi polynomials [34, 35]

\begin{eqnarray}
\psi_{n\ell}\,(z)=b_n\,P_n^{(a_3,\,\,
A)}\,(1-z)\,,\,\,\,\,\,A>-1\,,\,\,\,\,\,a_3>-1\,.
\end{eqnarray}
On the other hand, the other part of the wave function is obtained
from Eq. (16) as

\begin{eqnarray}
\xi(z)=z^{a_3/2}\,(2-z)^{A/2}\,.
\end{eqnarray}
Thus, the total eigenfunctions take

\begin{eqnarray}
\phi_{n\ell}\,(z)=b'_n\,(2-z)^{A/2}z^{a_3/2}\,P_n^{(a_3,\,\,A)}\,(1-z)\,.
\end{eqnarray}
where $b'_n$ is the new normalization constant. It is obtained
from

\begin{eqnarray}
\frac{\beta}{4}\,\int_{0}^{1}\left|\phi_{n\ell}(z)\right|^2\Big(\,\frac{z-2}{z}\,\Big)dz=1\,.
\end{eqnarray}
To evaluate the integral, we use the following representation of
the Jacobi polynomials [35]

\begin{eqnarray}
P_n^{(\sigma,\,\varsigma)}(z)&=&\,\frac{\Gamma(n+\sigma+1)}{n!\Gamma(n+\sigma+\varsigma+1)}\nonumber\\
&\times&\sum_{m=0}^{n}\,\Bigg(\begin{array}{c}
  n \\
  r
\end{array}\Bigg)\,
\frac{\Gamma(n+\sigma+\varsigma+m+1)}{\Gamma(m+\sigma+1)}\,
\frac{\Gamma(n+\sigma+\varsigma+m+1)}{\Gamma(m+\sigma+1)}\,(\,\frac{z-1}{2})^m\,,
\end{eqnarray}
where $\Bigg(\begin{array}{c}
  n \\
  r
\end{array}\Bigg)=\frac{n!}{r!(n-r)!}=\frac{\Gamma(n+1)}{\Gamma(r+1)\Gamma(n-r+1)}$\,.
Hence, from Eq. (41), and with the help of Eq. (42), we get

\begin{eqnarray}
[g(n,m)\times g(r,s)]\Big(\,\frac{\beta}{4}\,\Big)
\left|b'_n\right|^2\int_{0}^{1}z^{m+s+a_3-1}\,(2-z)^{A+1}\,dz=1\,,
\end{eqnarray}
where $g(n,m)$, and $g(r,s)$ are two arbitrary functions of the
parameters $A$, and $a_3$, and given by

\begin{eqnarray}
g(n,m)&=&\,\frac{2^{-m}\,\Gamma(A+n+1)}{n!\Gamma(A+a_3+n+1)}\nonumber\\
&\times&\,\sum_{m=0}^{n}\,\Bigg(\begin{array}{c}
  n \\
  r
\end{array}\Bigg)\,
\frac{\Gamma(n+\sigma+\varsigma+m+1)}{\Gamma(m+\sigma+1)}\,
\frac{\Gamma(A+a_3+n+m+1)}{\Gamma(a_3+m+1)}\,(-1)^m\,,\nonumber\\
\end{eqnarray}
and

\begin{eqnarray}
g(r,s)=g(n,m)\,(n \rightarrow r; m \rightarrow s)\,.
\end{eqnarray}

The integral in Eq. (43) can be evaluated by using the following
integral representation of hypergeometric type function
$_2F_1(a,b;c;z)$ [36]

\begin{eqnarray}
_2F_1(a,b;c;z)=\,\frac{\Gamma(c)}{\Gamma(b)\Gamma(c-b)}\,\int_{0}^{1}
t^{b-1}\,(1-t)^{c-b-1}\,(1-tz)^{-a}\,dt\,,
\end{eqnarray}
by setting the variable $z \rightarrow\,\frac{z}{2}$, and taking
$c=1+b$\,,\,$z=1$, one gets

\begin{eqnarray}
\int_{0}^{1}\,t^{b-1}\,(2-t)^{-a}\,dt=\,\frac{\Gamma(b)\Gamma(1)}{2^a\,\Gamma(1+b)}\,
_2F_1\,(a,b;1+b;\,\frac{1}{2}\,)\,,
\end{eqnarray}

From last equation

\begin{eqnarray}
\int_{0}^{1}\,z^{m+s+a_3-1}\,(2-z)^{A+1}\,dz&=&\,\frac{\Gamma(m+s+a_3)\Gamma(1)}{2^a\,\Gamma(m+s+a_3+1)}\nonumber\\
&\times&_2F_1(-A-1,m+s+a_3;m+s+a_3+1;\,\frac{1}{2}\,)\,,
\end{eqnarray}
where we set $b=m+s+a_3$\,, and $a=-A-1$.

By using the following identities of hypergeometric type functions
[36]

\begin{eqnarray}
_2F_1\,(a,b;c;-1)&=&\,\frac{\Gamma\Big(\,\frac{1}{2}b+1\Big)\Gamma(b-a+1)}{\Gamma(b+1)
\Gamma(\Big(\,\frac{1}{2}b-a+1\Big)}\,,\,\,\,(a-b+c=1\,,\,\,b>0)\\
_2F_1\,(a,b;c;\frac{1}{2})&=&2^a\,_2F_1\,(a,c-b;c;-1)\,,
\end{eqnarray}
the function of $_2F_1\,(a,b;c;z)$ in Eq. (48) can be evaluated as

\begin{eqnarray}
_2F_1\,(-A-1,b;1+b;\frac{1}{2})=\,2^{-A-2}\,\sqrt{\pi}\,\frac{\Gamma(A+3)}{\Gamma\Big(\,\frac{5}{2}\,+A\Big)}\,,
\,\,\,(m+s+a_3-A=2)\,,
\end{eqnarray}

Finally, we get the normalization constant as

\begin{eqnarray}
\left|b'_n\right|^2=\,\frac{8}{\beta\sqrt{\pi}}\,\frac{\Gamma(m+s+a_3+1)\Gamma\Big(\,\frac{5}{2}\,+A\Big)}
{\Gamma(m+s+a_3)\Gamma(3+A)[g'(n,m) \times g(r,s)]}\,.
\end{eqnarray}
where

\begin{eqnarray}
g'(n,m)&=&\,\frac{2^{-m}\,\Gamma(A+n+1)}{n!\Gamma(A+a_3+n+1)}\nonumber\\
&\times&\,\sum_{m=0}^{n}\,\Bigg(\begin{array}{c}
  n \\
  r
\end{array}\Bigg)\,
\frac{\Gamma(n+\sigma+\varsigma+m+1)}{\Gamma(m+\sigma+1)}\,
\frac{\Gamma(A+a_3+n+m+1)}{\Gamma(a_3+m+1)}\,(-1)^{m+1}\,.
\end{eqnarray}

\section{Conclusion}
We have solved the radial part of the KG-equation for the modified
Woods-Saxon potential in the case of position dependent mass by
using a new approximation scheme to the centrifugal potential term
for any $\ell$ values. It is observed that the results obtained by
using the new scheme for the case of the constant mass are
consistent with the ones obtained in Ref. [29]. It is seen that
there is a linear relation between the energy eigenvalues and the
contributions coming from the dependence of the mass on spherical
coordinate. The energy spectra and the corresponding wave
functions are obtained by applying the NU-method. The
eigenfunctions can be expressed in terms of Jacobi polynomials in
the scheme of the new approximation to the centrifugal barrier in
the case of position dependent mass.

\section{Acknowledgments}
This research was partially supported by the Scientific and
Technical Research Council of Turkey.

\newpage

\begin{table}
\caption{\label{tab:special}The dependence of the bound states for
a system 'proton+nucleon with average  mass number $A=56$' on the
parameter $m_{1}$ in $MeV$ for $q=1$ by using the numerical values
$m_{p}=1.007825$ amu, $V_{0}=47.78$ MeV, $r_{0}=4.91623$ fm [37].}
\begin{ruledtabular}
\begin{tabular}{cccc}
$m_{1} (amu)$ & $n$ & $\ell$ & $ E_{n\ell}<0$\\ \hline
0 & 0 & 0 & 171.920\\
  & 1 & 0 & 922.962\\
  &   & 1 & 924.286\\
  & 2 & 0 & 891.947\\
  &   & 1 & 895.473\\
  &   & 2 & 902.084\\ \hline
0.01  & 0 & 0 & 270.028\\
  & 1 & 0 & 842.200\\
  &   & 1 & 846.735\\
  & 2 & 0 & 808.765\\
  &   & 1 & 813.490\\
  &   & 2 & 822.663\\ \hline
0.001  & 0 & 0 & 187.762\\
  & 1 & 0 & 915.806\\
  &   & 1 & 917.461\\
  & 2 & 0 & 844.123\\
  &   & 1 & 887.762\\
  &   & 2 & 894.605\\
\end{tabular}
\end{ruledtabular}
\end{table}

\end{document}